\documentclass[11pt]{article}
\usepackage{epsfig}
\usepackage{psfig}
\usepackage{cite}
\input{epsf}
\newcommand{\beq}{\begin{equation}}
\newcommand{\eeq}{\end{equation}}
\newcommand{\beqa}{\begin{eqnarray}}
\newcommand{\eeqa}{\end{eqnarray}}

\newcommand{\vs}{\vspace{-0.20cm}}
\begin{document}

%%\noindent Accepted for publication in Phys. Lett. {\bf B} 

\begin{flushright}
{\tiny{FZJ-IKP-TH-2006-07}}
{\tiny{HISKP-TH-06/11}}
\end{flushright}

\vspace{.6in}

\begin{center}

\bigskip

{{\Large\bf The nucleon axial--vector coupling beyond one loop}\footnote{This
research is part of the EU Integrated Infrastructure Initiative Hadron Physics Project
under contract number RII3-CT-2004-506078. Work supported in part by DFG (SFB/TR 16,
``Subnuclear Structure of Matter'').}}

\end{center}

\vspace{.3in}

\begin{center}
{\large 
V\'eronique Bernard$^\star$\footnote{email: bernard@lpt6.u-strasbg.fr} {\footnotesize and}
Ulf-G. Mei{\ss}ner$^\ddagger$$^\ast$\footnote{email: meissner@itkp.uni-bonn.de}
}

\vspace{1cm}

$^\star${\it Universit\'e Louis Pasteur, Laboratoire de Physique
            Th\'eorique\\ 3-5, rue de l'Universit\'e,
            F--67084 Strasbourg, France}

\bigskip

$^\ddagger${\it Universit\"at Bonn,
Helmholtz--Institut f\"ur Strahlen-- und Kernphysik (Theorie)\\
Nu{\ss}allee 14-16,
D-53115 Bonn, Germany}

\bigskip

$^\ast${\it Forschungszentrum J\"ulich, Institut f\"ur Kernphysik 
(Theorie)\\ D-52425 J\"ulich, Germany}

\bigskip

\bigskip

\end{center}

\vspace{.4in}

\thispagestyle{empty} 

\begin{abstract}\noindent 
We analyze the nucleon axial-vector coupling to two loops
in chiral perturbation theory.
We show that chiral extrapolations based on this representation
require lattice data with pion masses below 300 MeV.
\end{abstract}

\vfill

\pagebreak

%%%%%%%%%%%%%%%%%%%%%%%%%%%%%%%%%%%%%%%%%%%%%%%%%%%%%%%%%%%%%%%%%%%%%%%
\noindent {\bf 1.} 
The axial--vector coupling constant $g_A$ is a fundamental property
of the nucleon that can e.g. be determined in neutron $\beta$--decay
(for a review on the nucleons axial properties, see~\cite{Bernard:2001rs}).
It is directly related to the fundamental pion--nucleon coupling
constant by the Goldberger-Treiman relation and thus of great importance
for the problem of nuclear binding. In the last few years, first 
attempts to calculate $g_A$ using various approximations to QCD on 
a discretized space--time (lattice QCD) have been published, see 
e.g.~\cite{Ohta:2002ns,Sasaki:2003jh,Khan:2004vw,Edwards:2005ym}.
These results are obtained for quark masses considerably larger than
the physical ones, the lowest quark masses considered e.g. in the
most recent study~\cite{Edwards:2005ym} correspond to a pion mass of
about 350~MeV (which should already be close to the so--called chiral regime). 
It is therefore necessary to
perform a chiral extrapolation to connect these lattice results with
the physical values of the quark masses.\footnote{In addition, one
has to correct for finite volume and finite size effects, which we do 
not consider in what follows. For  recent studies, 
see~\cite{Beane:2004rf,Khan:2006de}.} 
Long before the advent of these lattice
data it was noted that the chiral expansion of 
the axial--vector coupling does not show the
expected convergence behaviour for an SU(2) quantity --- the correction
of order $M_\pi^3$ is of the order of 30\% at the physical pion
mass although it is two orders down
compared to the leading term~\cite{Kambor:1998pi}. One therefore
can not expect the one--loop representation to be very accurate
for increasing pion mass. In fact, the complete (fourth order) 
one--loop result is steeply
rising with growing $M_\pi$ while the lattice data show essentially
no pion mass dependence~\footnote{One should be somewhat cautious to
draw too strong conclusions from such observations
because most of the available lattice QCD results are far outside
the range of applicability of chiral perturbation theory or any 
model-independent scheme.}. A possible solution to this problem was
offered in Ref.~\cite{Hemmert:2003cb} where an effective field theory
with explicit delta degrees of freedom at leading one--loop order
could lead to a flat pion mass dependence of $g_A$,
requiring, however, a  fine tuning of certain low--energy constants.
For a recent update, see~ \cite{Procura:2005ev}.
Two remarks on that result are in order: First, it should also
be noted that most lattice data available at that time are far outside
the regime of applicability of the effective field theory. Second,
to judge upon the usefulness of such an approach 
requires a systematic analysis of many other observables
which has not been done so far\footnote{This was to some extent attempted in
 \cite{Hemmert:2003cb} where one LEC was constrained by matching to the
pion-nucleon theory, in which this LEC had been determined earlier.} 
(for further discussions on this issue,
see e.g.~\cite{Meissner:2005ba,Colangelo:2005cg}). To obtain a deeper 
understanding of the chiral expansion of $g_A$ it appears therefore timely
to go beyond the one--loop approximation in nucleon chiral perturbation theory.
This is precisely the issue of this
letter. Using renormalizations group methods, we will determine 
the coefficient of the double log in $g_A$ -- that arises first at two--loop 
order  -- from the existing one--loop results (note that the spectral
function of the axial form factor to two loops was already worked out
in Ref.~\cite{Bernard:1996cc}). We also give the general structure 
of the two--loop representation of $g_A$, and  determine the numerically
leading contributions to the single logarithm and polynomial terms at order
$M_\pi^4$ and $M_\pi^5$, generated by graphs with insertion proportional to
the large dimension two low--energy constants $c_3$ and $c_4$ and the 
dimension three LEC $\bar{d}_{16}$. We thus achieve  more detailed 
information  on the quark mass expansion of the nucleon axial--vector 
coupling constant.

\medskip

%%%%%%%%%%%%%%%%%%%%%%%%%%%%%%%%%%%%%%%%%%%%%%%%%%%%%%%%%%%%%%%%%%%%%%%%%%%%%%
\noindent {\bf 2.} 
Our calculation is based on the effective
Lagrangian of pions and nucleons coupled  to external sources.
The various contributions to S-matrix elements and transition 
currents are organized in powers of the small parameter $q$,
where $q$ collectively denotes small pion four-momenta, the
pion mass and baryon three-momenta. The effective SU(2) Lagrangian
is given as a string of terms with increasing chiral dimension,
\beq
{\cal L}_{\rm eff} = {\cal L}_{\pi N}^{(1)} +  {\cal L}_{\pi N}^{(2)} +
{\cal L}_{\pi N}^{(3)} +  {\cal L}_{\pi N}^{(4)} +
{\cal L}_{\pi N}^{(5)} +  {\cal L}_{\pi N}^{(6)} +
{\cal L}_{\pi \pi}^{(2)} +  {\cal L}_{\pi \pi}^{(4)} + \dots
\eeq
where the ellipsis denotes terms not needed in what follows. 
The local operators at the various orders are accompanied by
low-energy constants (LECs), these are denoted as $c_i, d_i, e_i,
\ldots$ for the dimension two, three, four, $\ldots$ 
pion-nucleon Lagrangian and $l_i$ for the mesonic LECs of
dimension four. According to the power counting, the tree
approximation is given by tree graphs with insertions from 
${\cal L}_{\pi N}^{(1,2)}$. The one--loop
approximation contains further tree graphs with insertions from 
${\cal L}_{\pi N}^{(3,4)}$ and one--loop graphs with
insertions from ${\cal L}_{\pi N}^{(1)}$ and at most one insertion
from ${\cal L}_{\pi N}^{(2)}$. At two--loop order, we have two--loop
graphs with insertions from ${\cal L}_{\pi N}^{(1)}$ and at most 
one insertion from ${\cal L}_{\pi N}^{(2)}$, one--loop graphs with
insertions from ${\cal L}_{\pi N}^{(3,4)}$ and further tree graphs
related to ${\cal L}_{\pi N}^{(5,6)}$ (and the corresponding mesonic
contributions). Since we are interested in the quark mass
expansion of the axial-vector coupling $g_A$, it is most convenient
to work in the heavy baryon framework (for a review, 
see~\cite{Bernard:1995dp}). In two--flavor chiral perturbation theory, 
the quark mass expansion is mapped onto an expansion in the pion mass,
whose physical value is denoted by $M_\pi$. Consequently, the
chiral expansion of $g_A$ takes the form
\beqa\label{gAstruc}
g_A &=& g_0 \,\, \biggl\{ 1 + \left( \frac{\alpha_2}{(4\pi F)^2} \ln
\frac{M_\pi}{\lambda} + \beta_2 \right) \, M_\pi^2 + \alpha_3 \, M_\pi^3
\nonumber\\
&& \quad + \left(\frac{\alpha_4}{(4\pi F)^4} \ln^2\frac{M_\pi}{\lambda}
+  \frac{\gamma_4}{(4\pi F)^2} \ln\frac{M_\pi}{\lambda} + \beta_4
\right) \, M_\pi^4 + \alpha_5 \, M_\pi^5 \biggr\} + {\cal O}(M_\pi^6)~,
\nonumber \\
&=&  g_0 \,\, \biggl\{ 1 + \Delta^{(2)} + \Delta^{(3)} + \Delta^{(4)} +
    \Delta^{(5)} \biggr\}   + {\cal O}(M_\pi^6)~,
\eeqa
with $g_0$ the chiral limit value of $g_A$, $g_A = g_0 [ 1 + {\cal O}
(M_\pi^2)]$, 
$\lambda$ is the scale of dimensional regularization, and $\Delta^{(n)}$
denotes the relative correction at order $M_\pi^n$.
Further, $F$ denotes the pion decay constant in the chiral limit,
$F_\pi = F [1 + {\cal O}(M_\pi^2)]$. To the order
we are working, we require the quark mass expansion of $F_\pi$,
\beq\label{Fpirenorm}
F_\pi  = F\, \left[ 1 + \frac{M^2_\pi}{16\pi^2 F^2} \bar{\ell}_4 
+ {\cal O}(M_\pi^4) \right]~, 
\eeq
in terms of  the scale--independent LEC 
\beq\label{eq:l4}
\bar{\ell}_4 = 16\pi^2 \ell_4^r (\lambda) - 2\ln (M_\pi/\lambda)~,
\eeq
where $\ell_4^r (\lambda)$ is the corresponding
scale--dependent renormalized LEC. Note that this explicit quark mass 
dependence of $F_\pi$ has to be accounted for when one studies the axial
coupling as a function of the pion mass. Note also that when we generate 
the numerical value of $ \ell_4^r (\lambda)$ from   $\bar\ell_4$, we have
of course to use the physical value of the pion mass in Eq.~(\ref{eq:l4}). 
Furthermore, the chiral expansion of the pion decay constant generates 
contributions to $\alpha_4$, $\beta_4$,
$\gamma_4$ and $\alpha_5$. This can be seen from Eq.~(\ref{gAstruc}) which
is expressed in terms of the chiral limit value $F$ instead of the physical
value, as it is commonly done. At a fixed pion mass, these two representations
are of course equivalent. The explicit expressions of these additional quark
mass dependent terms are given below. The third order one-loop coefficients 
$\alpha_2$ and $\beta_2$ in Eq.~(\ref{gAstruc}) were first given 
in~\cite{Bernard:1992qa} and the one--loop fourth order calculation 
was completed in ~\cite{Kambor:1998pi} with (we use the by now standard
notation of Refs.~\cite{Bernard:1995dp,Fettes:1998ud})
\beqa\label{oneloop}
\alpha_2 &=& -2 -4g_0^2~, \nonumber\\ 
\beta_2 &=& \frac{4}{g_0} \biggl({d}_{16}^r(\lambda) - 2 g_0 d^r_{28}(\lambda)
\biggr) - \frac{g_0^2}{(4\pi F)^2}~, 
\nonumber\\ 
\alpha_3 &=& \frac{1}{24\pi F^2 m_0}
\left(3+3g_0^2-4m_0c_3+8m_0c_4\right)~,
\eeqa
with 
\beqa\label{eq:d1628}
{d}_{16}^r(\lambda) &=& \bar{d}_{16} + \frac{g_0(4-g_0^2)}{8(4\pi F)^2}\ln
\frac{M_\pi}{\lambda}~, \nonumber\\
{d}_{28}^r(\lambda) &=& \bar{d}_{28} - \frac{9g_0}{16(4\pi F)^2}\ln
\frac{M_\pi}{\lambda}~.
\eeqa
As in \cite{Fettes:1998ud}, we set $\bar{d}_{28} = 0$ in what follows. 
Again, note that these relations for ${d}_{i}^r(\lambda)$ $(i = 16,28)$ 
have to be taken at the physical value of $M_\pi$ when it comes to pin
down their numerical values.
The dimension two LECs $c_3, c_4$ can be determined e.g. from 
the analysis of elastic
pion--nucleon scattering and the dimension three LEC $d_{16}$ from the
reaction $\pi N \to \pi\pi N$ (for a detailed discussion see e.g.
Ref.~\cite{Meissner:2005ba} and references therein).  
In Eq.~(\ref{oneloop}), $m_0$ denotes the
nucleon mass in the chiral limit. To the order we are working, it is
related to the physical nucleon mass $m_N$ via
\beq
m_N   =  m_0 - 4 c_1 M_\pi^2 + {\cal O}(M_\pi^3) ~,
\eeq
with $c_1$ another dimension two LEC that can be determined e.g. from
low energy pion-nucleon scattering data or the pion--nucleon sigma term. 
This quark mass dependence of
the nucleon mass induces  corrections at fifth order from the
third order coefficient $\alpha_3 \sim 1/m_N$. In what follows, we always
work with $m_0$ and absorb this induced contribution in the combination of
LECs contributing to $\alpha_5$. 

\smallskip\noindent
In this letter, we are going to evaluate
the coefficient $\alpha_4$ of the double logarithm which arises at
two--loop order. This requires only parameters from the one--loop
calculation, as already stressed by Weinberg in his seminal 
paper~\cite{Weinberg:1978kz}. The coefficients $\beta_4, \gamma_4$ and
$\alpha_5$ contain combinations of LECs from ${\cal L}_{\pi N}^{(2,3)}$ and
unknown LECs from ${\cal L}_{\pi N}^{(4,5,6)}$, we
will estimate these using naturalness arguments and also from the
description of the available lattice data (at small enough pion masses).
Note that we can, in addition, work out the numerically large contributions to
these coefficients $\sim c_i/m_0$ and $\sim c_i/m_0^2$ from the expansion
of the corresponding relativistic one-loop graphs (together with the induced
contributions from the quark mass expansion of $F_\pi$).

\medskip

\begin{figure}[tb]
\centerline{
\epsfig{file= 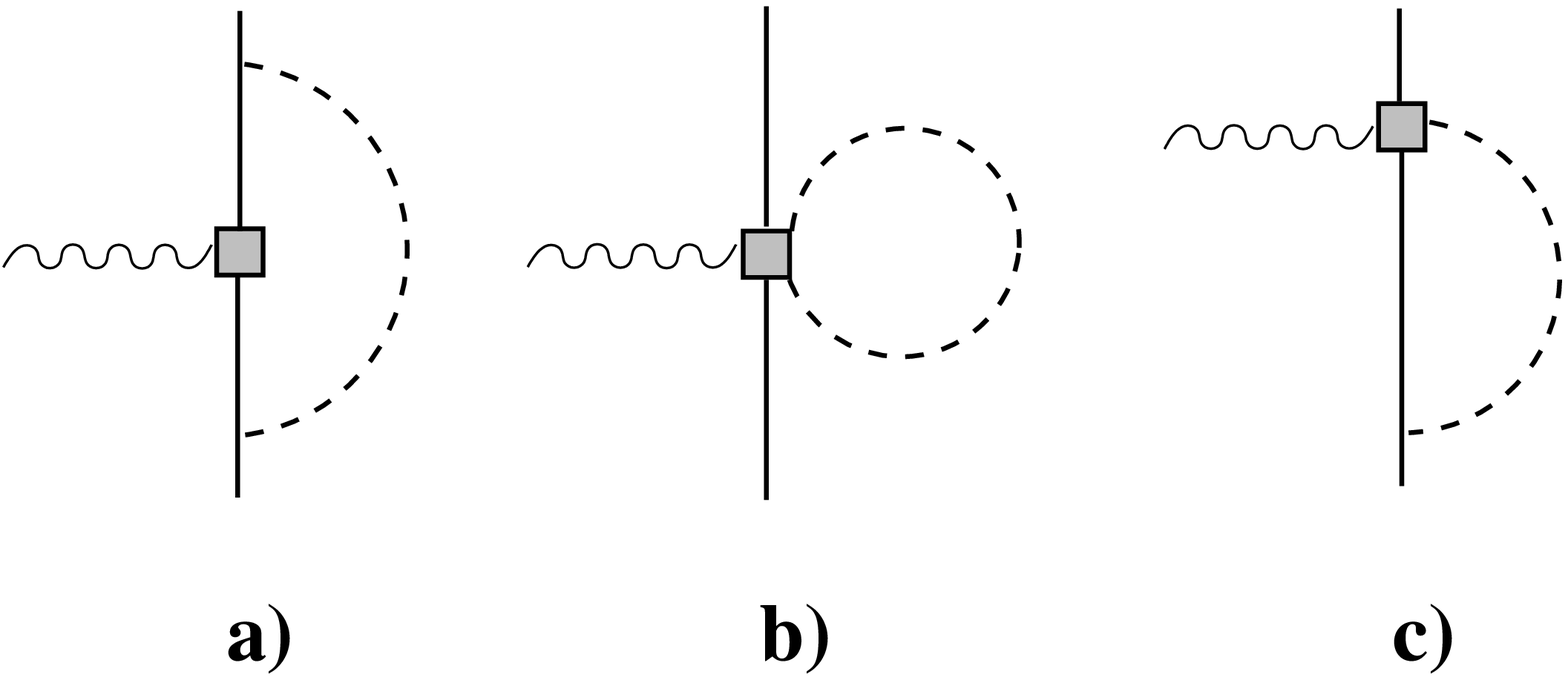,width=.95\textwidth,silent=,clip=}
}
\vspace{0.1cm}
\begin{center}
\caption{Topologies of the one--loop graphs that generate
the coefficient of the double log at two--loop order. The hatched square
denotes a dimension three insertion proportional to some of the LECs $d_i$.
\label{fig:dia}}
\vspace{-0.5cm}
\end{center}
\end{figure}

%%%%%%%%%%%%%%%%%%%%%%%%%%%%%%%%%%%%%%%%%%%%%%%%%%%%%%%%%%%%%%%%%%%%%%%%%%%%%%%%%%%%%%%%
\noindent {\bf 3.} 
The application of renormalization group (RG) methods to chiral effective
Lagrangians was pioneered by Weinberg~\cite{Weinberg:1978kz}. He showed that
the coefficient of the double log $\sim \ln^2 M_\pi$ can be entirely expressed
in terms of coupling constants of the one--loop generating functional.
For recent applications of such RG methods in chiral perturbation theory 
for mesons, see e.g.
\cite{Colangelo:1995jm,Bijnens:1998yu,Buchler:2005xn}, a nice
discussion of this and related issues is given in~\cite{Buchler:2003vw}. Here,
we wish to apply the same arguments to the effective pion--nucleon Lagrangian.
According to the power counting, the double logs are generated from two--loop
graphs at ${\cal O}(q^5)$. Employing a mass--independent renormalization
scheme (here: dimensional regularization), the two--loop divergences take 
the generic form
\beq\label{twoloopdiv}
k(d) \, \frac{\lambda^{2\epsilon}}{(4\pi)^4} \, \left[ \frac{1}{\epsilon^2} +
 \frac{2}{\epsilon} \, \ln \frac{M_\pi}{\lambda} + \ln^2 \frac{M_\pi}{\lambda}
+ \ldots \right]~,
\eeq
with $d$ the number of space--time dimensions, $\epsilon = d - 4$ and $k(d)$ is
a function of $d$ that depends on the specific diagram under
consideration. This function can also be expanded around $d=4$, $k(d) = k_0
+ k_1 \epsilon + {\cal O}(\epsilon^2)$. The leading term in this expansion
generates the non-local divergence $\sim k_0 \ln M /\epsilon$ that must be
cancelled by one--loop graphs with insertions from the dimension three
effective pion--nucleon Lagrangian (parameterized by the unrenormalized LECs $d_i$). 
Such graphs give the generic contribution
\beq\label{twoloop}
-\frac{h_i(d)}{2} \, \frac{\lambda^{2\epsilon}}{(4\pi)^4} \, \left[ 
 \frac{\kappa_i}{\epsilon^2} +  \frac{\kappa_i}{\epsilon} \ln
 \frac{M_\pi}{\lambda} +  \frac{(4\pi)^2 d_i^r (\lambda)}{\epsilon}
+ (4\pi)^2 \, d_i^r  (\lambda)  \ln  \frac{M_\pi}{\lambda} 
+ \ldots \right]~,
\eeq
where the $d_i$ are the dimension three LECs that have the
form~\cite{Fettes:1998ud}
\beq
d_i (d) = \lambda^\epsilon \, \left[ \frac{\kappa_i}{(4\pi)^2 \epsilon}
+ d_i^r (\lambda) + \ldots \right]~,
\eeq
where we use the basis of operators enumerated in~\cite{Fettes:1998ud} with the
corresponding  $\beta$--functions $\kappa_i$ listed there.  Here, $h_i(d)$ is
a function specific for the coefficient under consideration, that itself 
depends on $d$ via $h_i(d) = h_{i0} + h_{i1} \epsilon  + {\cal O}(\epsilon^2)$.
As shown by Weinberg, the elimination of the non--local divergence is
guaranteed by the RG condition
\beq
k_0 = \frac{1}{4} \, h_{i0} \, \kappa_i~.
\eeq  
Using this equation, we can now calculate the coefficient of the double
log. There are two types of diagrams contributing,
namely irreducible and reducible ones, the latter being related to wave
function renormalization. It is important to note that the notion of reducibility
here refers to the two-loop graphs. The non-vanishing irreducible two-loop
contribution is generated from the graphs  shown in
Fig.~\ref{fig:dia}. The following operators (given in terms of their LECs)
contribute to the various graphs: a) $d_{16}$, $d_{25}$ b) $d_{10}$, $d_{11}$,  
$d_{12}$, $d_{13}$, and $d_{16}$, c), $d_1$, $d_2$, $d_{14}$, $d_{26}$ and
$d_{30}$ d) $d_{16}$ and $d_{29}$,
e) $d_{24}$ and $d_{28}$, f) $d_{16}$,  $d_{25}$ and $d_{29}$, 
g) $d_{26}$, $d_{27}$
and $d_{28}$, and h) $d_{24}$ and $d_{28}$.
\noindent
Using the $\beta$-functions from~\cite{Fettes:1998ud} gives the
contribution to the double log generated by these diagrams. It reads
\beq\label{eq:inteom}
\alpha_4^{\rm irr} = 2 \, \left( \frac{4}{3} + \frac{5}{3}  g_0^2 - g_0^4\right)~.
\eeq
Furthermore, there are reducible graphs generated from wave function
renormalization. These are given by $g_A^{\rm 1-loop} \cdot Z^{\rm 1-loop}$
and their contribution to the double log is
\beq\label{eq:redu}
\alpha_4^{\rm red} = 9\,\left(  g_0^2 + 2g_0^4\right) ~.
\eeq
Putting pieces together (i.e. the contributions from the irreducible 
diagrams, Eq.~(\ref{eq:inteom}), the reducible graphs, Eq.~(\ref{eq:redu}), 
and the induced term $\tilde{\alpha}_4$, 
see Eq.~(\ref{eq:induced}) below), we have thus for the coefficient $k_0 \equiv \alpha_4$ 
of the double log
\beq\label{eq:inteomtot}
\alpha_4 = \alpha_4^{\rm irr} + \alpha_4^{\rm red} + \tilde\alpha_4 =
-\frac{16}{3} - \frac{11}{3} g_0^2 + 16 g_0^4~. 
\eeq
This is the central result of this paper and allows us to analyze the
leading two--loop correction to the axial--vector coupling constant.
In the formulation using the Lagrangian given in~\cite{Fettes:1998ud}, 
one has to deal with a large number of equation
of motion terms. These can be, however, eliminated from the effective
Lagrangian as done in \cite{Ecker:1995rk}. We have therefore also
performed the calculation in the basis given in that paper and
employing the pertinent $\beta$-functions. 
We find the same result as in Eq.(\ref{eq:inteomtot}),
which serves as an excellent check on our calculation\footnote{Provided one
corrects for the typographical error  in $\beta_{11}$ in that paper, 
see also \cite{Gasser:2002am}.}.
Note that this procedure generates also part of the
single log coefficient $\gamma_4$ in Eq.~(\ref{gAstruc}). We only
give the contribution generated from the operator proportional 
to the axial LEC $d_{16}$ (note that some of the other LECs $d_i$ are
also known, see Ref.~\cite{Fettes:1999wp}, but only $d_{16}$ plays 
a prominent role in the chiral expansion of $g_A$)
\beq\label{eq:d16}
\gamma_4^{d_{16}} =  -12\, d_{16}^r (\lambda ) \left( \frac{5}{3g_0} + g_0\right)~.
\eeq
Note further that using a relativistic formulation (see
e.g. Ref.~\cite{Schweizer}), it is easy to work out the $1/m_0$ and
$1/m^2_0$ corrections to the
large contribution $\alpha_3 \simeq 94$ (for the parameters given below,
see also  the discussion in \cite{Meissner:2005ba}) 
that will give a sizeable contribution to the coefficients 
$\beta_4, \gamma_4$ and $\alpha_5$, respectively.  These terms are given 
by
\beqa
\gamma_4^{c_i} &=& \frac{4(c_4-c_3)}{m_0}~, \nonumber\\
\beta_4^{c_i} &=& \frac{c_4}{m_0} \, \frac{1}{4\pi^2F^2}~, \nonumber\\
\alpha_5^{c_i} &=&  \frac{c_3}{m_0^2} \, \frac{1}{16\pi F^2} ~. 
\eeqa
The numerical values of these contributions will be given below, but we
remark already that in particular $\gamma_4^{c_i}$ will contribute sizeably.
Finally, we collect here the induced terms form the quark mass expansion
of the pion decay constant, cf. Eq.~(\ref{Fpirenorm}). Denoting these by a
tilde, they read
\beqa\label{eq:induced}
\tilde{\alpha}_4 &=& 4 \, \alpha_2~, \nonumber\\ 
\tilde{\gamma}_4 &=&  -\frac{2}{F^2} \, \alpha_2 \, l_4^r (\lambda) 
- \frac{4g_0^2}{(4\pi F)^2}~, \nonumber\\ 
\tilde{\beta}_4 &=& \frac{2g_0^2}{(4\pi F)^2 F^2} l_4^r (\lambda)~, \nonumber\\
\tilde{\alpha}_5 &=& -\frac{2 \alpha_3}{F^2}\left( l_4^r (\lambda)
- \frac{1}{8 \pi^2} \ln \frac{M_\pi}{\lambda} \right)~.
\eeqa 
We end this section with a brief comment of the one-loop chiral EFT
representation given in~\cite{Hemmert:2003cb}. Including an explicit 
delta to leading one--loop order generates some of the terms $\sim M_\pi^3$ 
and  part of the coefficients $\gamma_4$ and $\beta_4$ (plus some
other higher order terms $\sim M_\pi^{2n}/\Delta^{2m}$ with $2n-2m =2$ and
$\Delta$ is the delta-nucleon mass splitting). However, already at 
third order in the pion mass, 
this neglects other resonance contributions to the LECs $c_3$ and $c_4$  
(for a detailed discussion, see e.g.~\cite{Bernard:1996gq})
and therefore that representation can not be considered as accurate as the
one developed here (for pion masses in the chiral regime). 

\medskip

%%%%%%%%%%%%%%%%%%%%%%%%%%%%%%%%%%%%%%%%%%%%%%%%%%%%%%%%%%%%%%%%%%%%%%%%%%%
\noindent {\bf 4.}
We are now in the position to put pieces together.
First, we consider the contribution of the various (incomplete)
terms at fourth and fifth order in the pion mass for the physical
values of the quark masses. For that, we express all parameters in
terms of their chiral limit values. We take $\bar\ell_4 = 4.33$ 
corresponding to $F = 87\,$MeV and $m_0 = 880\,$MeV. If not stated
otherwise, we use $\bar{d}_{16} = -1.76\,$GeV$^{-2}$, $c_3 = 3.5\,$GeV$^{-1}$ and
$c_4 = -4.7\,$GeV$^{-1}$. We also work at $\lambda = m_0$. Note that we have
varied these LECs within their allowed ranges, but this did not lead to any
sizeable changes to the results given below. Furthermore,
we vary $g_0$ between 1.0 and 1.2. To be definite, let us set $g_0 =1$.
Collecting  pieces, we obtain
\beqa\label{eq:values}
\alpha_4 &=& \alpha_4^{{\rm irr + red}} + \tilde{\alpha}_4 = 31-24 = 7~,
\nonumber\\ 
\gamma_4 &=& \gamma_4^{c_i} + \tilde{\gamma}_4 + \gamma_4^{d_{16}} 
= (37.3 + 3.2 + 74.8)~{\rm GeV}^{-2} = 115.3~{\rm GeV}^{-2}~,
\nonumber\\ 
\beta_4 &=& \beta_4^{c_i} + \tilde{\beta}_4 = (13.3 + 0.9)~{\rm GeV}^{-4} 
= 14.2~{\rm GeV}^{-4}~,
\nonumber\\ 
\alpha_5 &=& \alpha_5^{c_i} + \tilde{\alpha}_5 =  (-16.0 -4.3)~{\rm GeV}^{-5} 
= -20.3~{\rm GeV}^{-5}~.
\eeqa
Of course, the coefficients $\gamma_4$, $\beta_4$ and $\alpha_5$ receive
further corrections from LECs that have to be determined e.g. from an 
analysis of lattice data or estimated assuming naturalness. 
We denote these additional contributions by  $\gamma_4^f$, $\beta_4^f$ and
$\alpha_5^f$. For the moment, we set $\gamma_4^f = \beta_4^f = \alpha_5^f = 0$.
In the notation of Eq.~(\ref{gAstruc}), these results translate into
\beqa\label{eq:delta}
\Delta^{(2)} &=& -15.3~\%~, \nonumber\\
\Delta^{(3)} &=&  25.6~\%~, \nonumber \\
\Delta^{(4)} &=&  \Delta^{(4)}_\alpha + \Delta^{(4)}_\gamma +
\Delta^{(4)}_\beta = (0.6 - 6.3 + 0.5)~\% = -5.6~\%~, \nonumber\\
\Delta^{(5)} &=& -0.1~\%~.
\eeqa
The sizeable fourth order contribution is entirely due to the large
coefficient $\gamma_4$, largely due to the insertion of the operator
$\sim d_{16}$, cf. Eq.~(\ref{eq:d16}). Note also that the fifth order term is very
small at the physical point. Ignoring the higher order corrections,
one can calculate the chiral limit value of $g_A$ from Eq.~(\ref{gAstruc})
using the values collected in Eq.~(\ref{eq:delta}),
\beq
g_0 = 1.21 ~[1.12]~,
\eeq
where the number in the brackets refers to the choice $\bar{d}_{16} = 
-0.92\,$GeV$^{-2}$ and we use $g_A = 1.267$.  Of course, these numbers
will be affected by the unknown LEC contributions $\gamma_4^f$, $\beta_4^f$
and $\alpha_5^f$. These values are consistent with the findings
in~\cite{Khan:2006de}. We also remark that the chiral expansion of $m_N$ is
much better behaved and one thus can successfully apply one--loop
extrapolation functions to pion masses below 450~MeV (for detailed discussions,
see e.g.~\cite{Bernard:2003rp,Frink:2005ru,Meissner:2005ba}). 

\smallskip
\noindent

\begin{figure}[tb]
\centerline{
\epsfig{file= 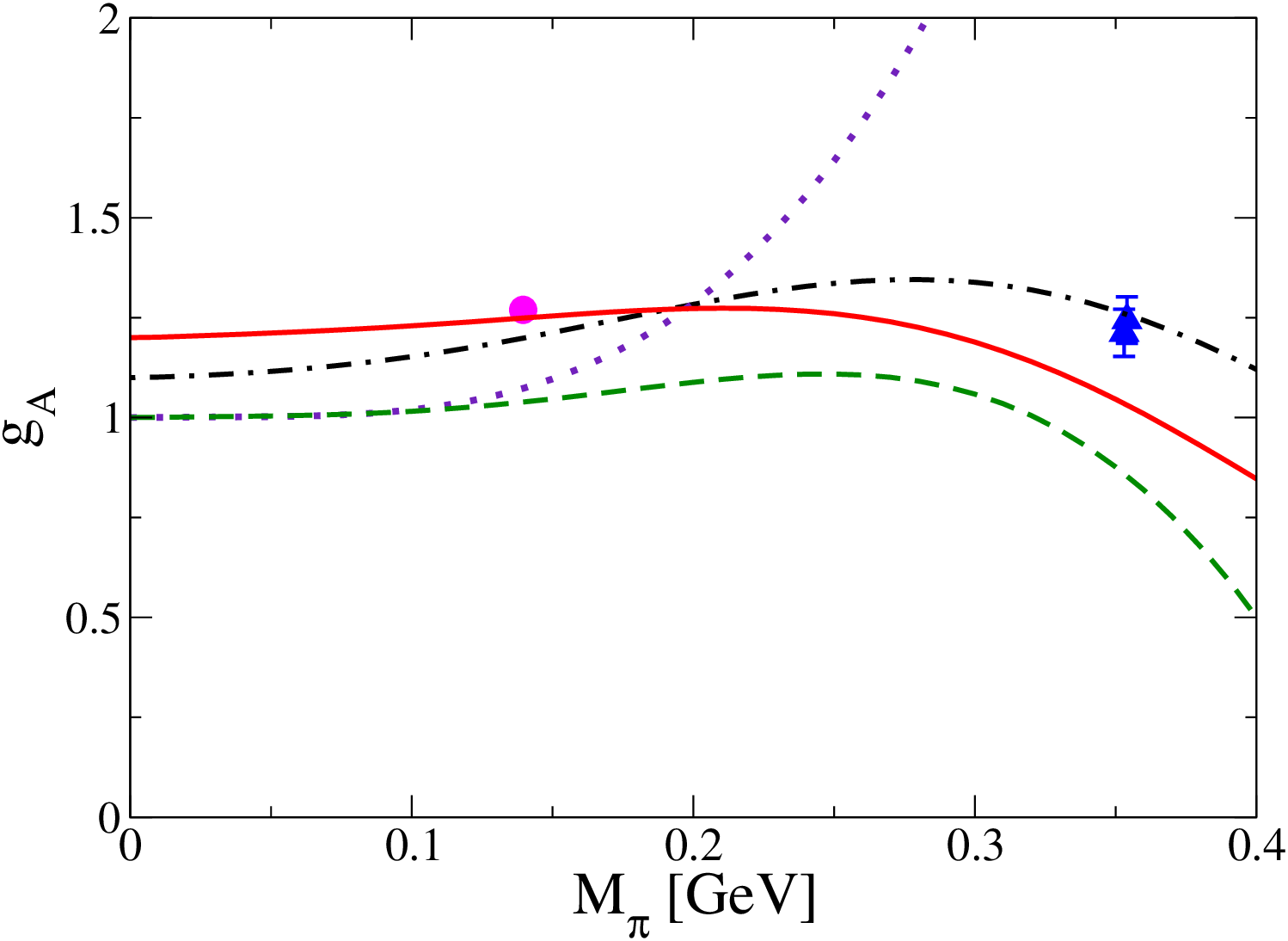,width=.55\textwidth,silent=,clip=}
}
\vspace{0.1cm}
\begin{center}
\caption{The axial-vector coupling as a function of the pion mass.
Solid (red) line: $g_0 = 1.2, \bar{d}_{16} = -1.76\,$GeV$^{-2}$, $\gamma_4^f =
50\,$GeV$^{-2}$, $\beta_4^f = 60\,$GeV$^{-4}$, $\alpha_5^f = 20\,$GeV$^{-5}$;
Dot-dashed (black) line: $g_0 = 1.1, \bar{d}_{16} = -0.92\,$GeV$^{-2}$, $\gamma_4^f =
40\,$GeV$^{-2}$, $\beta_4^f = 20\,$GeV$^{-4}$, $\alpha_5^f = 50\,$GeV$^{-5}$;
Dashed (green) line: $g_0 = 1.0, \bar{d}_{16} = -1.76\,$GeV$^{-2}$, $\gamma_4^f =
-50\,$GeV$^{-2}$, $\beta_4^f = \alpha_5^f = 0$. The dotted (violet) line is the
complete one-loop result with $g_0 = 1, \bar{d}_{16} = -1.76\,$GeV$^{-2}$ and using the
physical values of the nucleon mass and the pion decay constant.
The (magenta) circle denotes
the physical value of $g_A$ at the physical pion mass and the triangles are 
the lowest mass data from 
Ref.~\protect\cite{Edwards:2005ym}.
\label{fig:gA}}
\vspace{-0.7cm}
\end{center}
\end{figure}
\noindent
We show in Fig.~\ref{fig:gA} some typical examples for the
pion mass dependence of $g_A$ for values of $\gamma_4^f, \beta_4^f,
\alpha_5^f$ that lead to an approximately flat behaviour for not too high
pion masses. These values are of  natural size as a comparison with the induced 
pieces collected in Eq.~(\ref{eq:values}) reveals. This is very different
from the one--loop representation, which fails to generate a flat quark mass
dependence for values above the physical pion mass, 
see e.g.~\cite{Hemmert:2003cb,Meissner:2005ba}. We have studied many
more combinations of the LECs and found that for this representation to
be useful (that is leading to a moderate theoretical uncertainty), the pion 
mass should be less then 300~MeV. This can also been seen if one compares
to the complete one-loop result as depicted by the dotted line in 
Fig.~\ref{fig:gA}. The existing lattice results are at still
too high pion masses for a model-independent extrapolation to the physical
values of the quark masses. For a particular choice of the LECs, we can describe
the trend of the lattice data up to $M_\pi \simeq 600\,$MeV, but the theoretical
uncertainty is simply too large for such values of the pion mass.

\medskip

%%%%%%%%%%%%%%%%%%%%%%%%%%%%%%%%%%%%%%%%%%%%%%%%%%%%%%%%%%%%%%%%%%%%%%%%%
\noindent {\bf 5.} In this letter, we have studied the pion mass dependence
of the nucleon axial-vector coupling constant $g_A$. This is a fundamental
observable for our understanding of the nucleon structure in the regime of
strong QCD. First lattice simulations have appeared and so far, chiral
extrapolation functions appearing in the literature are either based on
(leading) one-loop chiral effective field theory results with explicit deltas
(for a critical discussion, see e.g. \cite{Colangelo:2005cg}) or are very 
model-dependent. We have provided the two--loop representation
in baryon chiral perturbation theory,
see Eq.~(\ref{gAstruc}), and determined the coefficient of the double log term
$\sim M_\pi^4 \ln^2 M_\pi$ based on renormalization group arguments. 
We have also determined some numerically important 
contributions to the terms $\sim M_\pi^4 \ln M_\pi, M_\pi^4$ and $M_\pi^5$.
We have shown that with LECs of natural size one can indeed obtain a flat
pion mass dependence of $g_A$ for pion masses below 400~MeV. 
We conclude that
lattice data for pion masses below 300~MeV are required to use this representation
with a moderate theoretical uncertainty. Such data should be available in the
near future.

\bigskip\bigskip

\noindent{\large {\bf Acknowledgements}}

\smallskip\noindent
We are grateful to Bugra Borasoy for some checks and Philipp H\"agler 
for providing us with the lattice data. We thank Gerhard Ecker for a
useful communication.

%%%%%%%%%%%%%%%%%% REFERENCES %%%%%%%%%%%%%%%%%%%%%%%%%%%%

\end{document}